# Electrically tunable ultrafast dynamics and interactions of hybrid excitons in a 2D semiconductor bilayer

Edoardo Lopriore[1,2*], Charalambos Louca[3,4*], Armando Genco[3§], Irantzu Landa[3], Daniel Erkensten[5], Charles J. Sayers[3], Samuel Brem[5], Raul Perea-Causin[6], Kenji Watanabe[7], Takashi Taniguchi[8], Christoph Gadermaier[3], Ermin Malic[5§], Giulio Cerullo[3,9], Stefano Dal Conte[3§], Andras Kis[1,2§]

[1]*Institute of Electrical and Microengineering, École Polytechnique Fédérale de Lausanne (EPFL), CH-1015 Lausanne, Switzerland*
[2]*Institute of Materials Science and Engineering, École Polytechnique Fédérale de Lausanne (EPFL), CH-1015 Lausanne, Switzerland*
[3]*Dipartimento di Fisica, Politecnico di Milano, Piazza Leonardo Da Vinci 32, 20133 Milano, Italy*
[4]*NanoPhotonics Centre, Cavendish Laboratory, Department of Physics, JJ Thompson Ave, University of Cambridge, Cambridge, UK*
[5]*Philipps-Universität Marburg, DE-35037, Marburg, Germany*
[6]*Stockholm University, SE-10691, Stockholm, Sweden*
[7]*Research Center for Electronic and Optical Materials, National Institute for Materials Science, 1-1 Namiki, Tsukuba 305-0044, Japan*
[8]*Research Center for Materials Nanoarchitectonics, National Institute for Materials Science, 1-1 Namiki, Tsukuba 305-0044, Japan*
[9]*CNR-IFN, Piazza Leonardo da Vinci 32, Milano, 20133, Italy*

*\* These authors contributed equally to this work.*
[§]*Correspondence should be addressed to: Andras Kis (andras.kis@epfl.ch), Stefano Dal Conte (stefano.dalconte@polimi.it), Ermin Malic (ermin.malic@uni-marburg.de, and Armando Genco (armando.genco@polimi.it)*

**ABSTRACT**

Extended efforts have been devoted to the study of strongly-interacting excitons and their dynamics, towards macroscopic quantum states of matter such as Bose-Einstein condensates of excitons and polaritons. Momentum-direct layer-hybridized excitons in transition metal dichalcogenides have attracted considerable attention due to their high oscillator strength and dipolar nature. However, the tunability of their interactions and dynamics remains unexplored. Here, we achieve an unprecedented control over the nonlinear properties of dipolar layer-hybridized excitons in an electrically gated van der Waals homobilayer monitored by transient optical spectroscopy. By applying a vertical electric field, we reveal strong Coulomb interactions of dipolar hybrid excitons, leading to opposite density-dependent energy shifts of the two main hybrid species based on their dipolar orientation, together with a strongly enhanced optical saturation of their absorption. Furthermore, by electrically tuning the interlayer tunneling between the hybridized carriers, we significantly extend the formation time of hybrid excitons, while simultaneously increasing their decay times. Our findings have implications for the search on quantum blockade and condensation of excitons and dipolaritons in two-dimensional materials.

**INTRODUCTION**

Light-matter interactions in van der Waals heterostructures of two-dimensional materials have been widely investigated for fundamental studies as well as optoelectronic device applications[1–3]. Linear and nonlinear optical properties of excitons in bilayers of transition metal dichalcogenides (TMDCs) can be tailored with unprecedented flexibility, allowing the exploration of macroscopic quantum states of matter[4,5] and novel quantum photonics applications[6]. Type-II TMDC heterobilayers can be optically excited to form strongly-bound inter-layer excitons with a characteristic out-of-plane static dipole due to the spatial separation of the electron and hole wavefunctions in different constituent layers[7–9]. Interlayer excitons are characterized by the field tunability of their emission energy via the quantum-confined Stark effect[10,11], with long lifetimes and diffusion lengths[12,13].

Evidence supporting the occurrence of high temperature Bose-Einstein condensation (BEC) of interlayer excitons in TMDC heterobilayers have been recently reported, although a clear observation of quantum phase transitions in these systems is still missing[14]. In the past, dipolar excitons have been strongly coupled to cavity photons in bulk double quantum wells, forming electrically-tunable dipolar polaritons[15], with enhanced nonlinear interactions towards high temperature polariton BEC[16]. However, the low oscillator strength of interlayer excitons in type-II van der Waals heterobilayers hinders the creation of interlayer polariton states.

On the other hand, layer-hybridized excitons (hIXs) in TMDC bilayers are characterized by an electron (or hole) that is delocalized between the two constituent layers through coherent tunneling, thus exhibiting a coexistence of inter- and intra-layer characters[18]. Their electrical tunability[19–23], allowing the modulation of their dynamics[24,25] and interactions, has been recently exploited to achieve anomalous diffusion regimes of transport with high exciton mobilities[24,26]. In particular, previous works on tunable momentum-indirect hIXs have focused on their power-independent quantum yield and repulsive interactions by their emission via phonon replicas[21,24]. However, the study of their ultrafast formation and interactions, as well as their strong coupling with light, are hindered by their momentum-indirect nature. On the other hand, momentum-direct hybrid excitons exhibit large oscillator strengths, allowing the observation of dipolar layer-hybridized exciton-polaritons (dipolaritons) with highly nonlinear effects both in homobilayers[27–29] and in heterobilayers[30,31]. Momentum-direct hIXs in bilayer 2H-$MoS_2$ have been proposed as optical probes for correlated many-body phases, such as interlayer exciton coherence[32], holding promise for the high-temperature superfluidity of hybrid exciton condensates[33]. Furthermore, bilayer TMDCs hosting hybrid species have been identified as an ideal platform for the realization of Bose-Einstein condensates of dipolar exciton ensembles[34]. Combining interlayer charge tunnelling with the twist angle degree of freedom has also allowed the realization of moiré-hybridized interlayer excitonic states, owing to the additional electronic band folding induced by the moiré superlattice[35,36]. Such hybrid moiré excitons have shown enhanced nonlinear interactions in the few-polaritons regime due to localization effects in the moiré potential, potentially useful for polariton quantum blockade[28].

For these reasons, the control and enhancement of momentum-direct hIX interactions, as well as their dynamics, is crucial to obtain macroscopic states of matter based on hybrid excitons and dipolaritons[37]. However, the field-dependent tunability of the interactions and the dynamics of momentum-direct hIXs remain unexplored. Moreover, while the formation dynamics of interlayer excitons in type-II heterobilayers have been recently investigated[38–40], the electrical tunability of the formation time for any interlayer excitonic species has not been reported until now.

Here, we optically investigate the density-dependent repulsive and attractive nonlinear interactions between differently oriented dipolar hybrid excitons in a MoS$_2$ homobilayer under an external electric field. We observe opposite energy shifts and a strongly enhanced optical saturation of exciton absorption compared to the zero-field case. Using transient reflectance spectroscopy, we demonstrate the tunability of layer-hybridized dipolar exciton interactions on ultrafast timescales. Finally, we electrically modulate the ultrafast formation and relaxation dynamics of momentum-direct hIXs, achieving their strong enhancement at a finite electric field. Our experiments are supported by a microscopic many-particle theory based on an equation-of-motion approach applied to excitons in TMD bilayers[41].

## RESULTS

**Layer-hybridized exciton species in MoS$_2$ homobilayers**

Our device consists of a fully hBN-encapsulated naturally stacked MoS$_2$ homobilayer with bottom and top graphene gates (Figure 1a, Supplementary Note 1). The dual-gate configuration allows us to apply a vertical electric field without inducing extrinsic electrostatic doping[11,24]. At zero applied vertical electric field, MoS$_2$ homobilayers host A and B excitonic resonances ($X_A$ and $X_B$) with predominantly intra-layer character, and an inter-layer excitonic transition that is strongly hybridized with $X_B$ due to a significant hole tunneling strength at the K/K' valleys ($t_v \approx 40$ meV)[42]. This hybrid species (hIXs) is characterized by a layer-localized electron and a delocalized hole wavefunction that is shared between both constituent MoS$_2$ layers.

The application of a vertical electric field lifts the degeneracy of hIXs due to the presence of two distinct interlayer dipolar orientations. The non-negligible out-of-plane static dipoles cause the two (KK) hIX configurations to display progressive shifts of their excitonic transitions in opposite energy directions. Here, we label the lower and higher-lying excitonic species as L-hIX and H-hIX, respectively, with tunneling rates $J_L$ and $J_H$ (Figure 1b). Consequently, as the electric field increases, L-hIX becomes more interlayer-like while H-hIX becomes more intralayer-like, as obtained from microscopic theory in Supplementary Figure 3.

To experimentally visualize the Stark effect of hybrid excitons in bilayer $MoS_2$, we illuminate our sample (Figure 1c) with broadband white-light supercontinuum pulses, and measure the reflectance contrast (RC) spectra as described in the Methods section. Figure 1d shows the RC spectra as a function of the applied vertical electric field. The dipolar nature of layer-hybridized species is evidenced by their Stark effect[22,24]. We extract from it an upper-bound estimate of ~0.2 nm for the effective dipole length of both L-hIXs and H-hIXs at near-zero electric fields (Supplementary Note 3). The measured field-dependent Stark shift magnitude is lower than the value extracted from the calculated energy landscape in Supplementary Note 2, but remains in good qualitative agreement. We attribute this discrepancy to intrinsic doping in the active area, which affects the field-dependent behaviour of dipolar excitonic species[24,43]. Evidence of intrinsic doping is also provided by the non-negligible trion peak $X_A^*$ (Figure 1e), which becomes visible at higher electric fields, probably due to electrostatic charge migration, whose characterization is beyond the scope of this work (Supplementary Note 3).

Figure 1e displays two characteristic RC spectra obtained at zero and high (230 mV/nm) electric fields. Even with a significant Stark shift, represented by a H-L energy splitting of $\delta E_{H-L} \simeq 40$ meV, the total oscillator strength of the two species remains approximately 20% of that of $X_A$, in agreement with the previous literature[44].

**Strong hIX nonlinearities due to dipolar interactions**

To investigate dipolar interactions (Figure 2), we illuminate the sample with single ultrashort (100 fs) white-light supercontinuum pulses of varying spectral bandwidths at a fixed vertical electric field (230 mV/nm) such that $\delta E_{H-L} \simeq 40$ meV. With this splitting, the L and H lineshapes can be well distinguished, given their linewidths in the range of 15 meV, while also being unaffected by the $X_A$ and $X_B$ tails. We first use a narrowband (NB) excitation configuration, where the pulses are spectrally filtered to excite only the hIX transitions (Methods) with a full-width half-maximum (FWHM) of 100 meV. As the pulse fluence is increased, we observe a progressive bleaching of the hIX peaks, together with a blueshift of the L-hIX peak energy and a redshift of the H-hIX peak energy (Figure 2a). The hIX density was determined through an experimental method that takes into account a convolution of the RC spectra and the laser profile. We note that these fluences correspond to hybrid exciton densities ($< 10^{12}$ cm$^{-2}$) well below the exciton Mott transition ($\sim 10^{14}$ $cm^{-2}$).

To understand the interactions at play, we developed a microscopic theory based on a hybrid exciton-exciton interaction Hamiltonian focusing on the dipolar interactions between hIXs (Supplementary Note 4). The density-dependent energy shift of hIXs is written as:

$$\Delta E^{i} = g_{x-x}^{i-i} n_i + g_{x-x}^{i-\bar{i}} n_{\bar{i}} \qquad (1)$$

where $i = H, L$ ($\bar{i}$ indicates the opposite species, i.e. if $i = H$, then $\bar{i} = L$ and vice versa), $n_i$ is the exciton density of a specific population, and $g_{x-x}$ is the dipolar hybrid exciton-exciton interaction. The interaction strength is determined by the interlayer mixing coefficient $C_{IX}$, which is effectively modulated by an applied vertical electric field (Supplementary Figure 3). We refer to $g_{x-x}^{i-i} \propto |C_{i,IX}|^2 |C_{i,IX}|^2 \, d_{TMD}/\epsilon_\perp$ and $g_{x-x}^{i-\bar{i}} \propto -|C_{i,IX}|^2 |C_{\bar{i},IX}|^2 \, d_{TMD}/\epsilon_\perp$ as the co-hIX and cross-hIX contributions, with repulsive and attractive characters, respectively. Here,

$d_{TMD}$ is the TMDC layer thickness, corresponding to the bare dipole, and $\epsilon_\perp$ is the out-of-plane component of the dielectric tensor of the TMDC. The H-L splitting at a fixed electric field is density dependent, as $\delta E(n_L, n_H) = \delta E_0 + \Delta E^L(n_L) + \Delta E^H(n_H)$. At vanishing fields, the interlayer composition of H and L is equal to $|C_{0,IX}|^2 \simeq 0.7$. We perform our calculations starting from a low-density splitting of $\delta E_0 \approx \delta E_{H-L}(n_{L,H} < 10^{10} \text{ cm}^{-2}) \simeq 40$ meV to best represent our experimental conditions. In this case, the calculated mixing coefficients amount to approximately $|C_{H,IX}|^2 \simeq 0.5$ and $|C_{L,IX}|^2 \simeq 0.8$, with decreasing and increasing interlayer character, respectively.

As a result, $\Delta E^L$ is dominated by repulsive interactions ($g_{x-x}^{L-L}$), giving rise to a density-dependent blueshift (Supplementary Note 2). Instead, given the weaker co-hIX interactions for H-hIXs, the attractive interactions between H-hIXs and L-hIXs ($g_{x-x}^{H-L}$) result in a density-dependent $\Delta E^H$ redshift. In Figure 2c, we present the measured hIX peak energy shifts $\Delta E^{L,H}$ extracted from the Lorentzian fits to the RC spectra, showing a good agreement with our theoretical calculations (dashed lines in Figure 2c). The theoretical shifts were taken by assuming $n_L \approx 2\, n_H$ based on a best-fit approach. As discussed in Supplementary Note 4, the density-dependent L-hIX blue-shift and H-hIX red-shift are qualitatively obtained regardless of the density ratio $n_L/n_H$, provided that $n_H \leq n_L$. This is justified by the fact that L-hIX is the lowest excited state among the hIX branches, and is aligned with their photoluminescence reported in the previous literature[45]. We note that many-body effects due to carrier exchange and dynamic screening are not included in the calculations, but are expected to modify quantitatively the density-dependent energy shifts[46,47]. In particular, it was shown recently that ultrafast dynamical screening strongly reduces the density-dependent energy shifts of interlayer excitons[46]. Since the measured energy shifts for hybrid excitons are larger than those for interlayer excitons, this may suggest that dynamical screening plays a smaller role in homobilayers. However, further experimental and theoretical studies are needed to confirm this interpretation.

We repeated our experiments in a broadband (BB) configuration, with the white-light supercontinuum pulses filtered to excite hIXs as well as $X_A$ (FWHM = 200 meV). In Figures 2d-e, we show the density-dependent quenching of the integrated RC for L-hIXs and H-hIXs peaks in both configurations. An exponential decrease of integrated RC for both L and H species is observed in the narrowband case for densities $n_{hIX} > 2 \cdot 10^{11}$ cm$^{-2}$. Supplementary Figure 6 further shows the bleaching and energy shift of hIXs with respect to illumination fluence in the BB case. Enhanced energy shifts and a stronger density-dependent optical saturation are found in the BB case due to inter-species interactions involving $X_A$ and hIXs. Under BB excitation, both $X_A$ and hIX are excited simultaneously, leading to additional interspecies interactions, as previously observed in the literature[27]. Notably, $X_A$ and hIX share holes in the same valence band via hole tunnelling, enabling a combination of phase space filling and inter-excitonic exchange interactions. Such tunnelling-enabled nonlinearity – activated only when the common valence band is jointly occupied under BB illumination – substantially enhances the bleaching of the hIX oscillator strength. Thus, in Figures 2d-e we observe a strong increase in nonlinearity for both L-hIX and H-hIX under the application of an electric field with BB excitation, as the reduction of their oscillator strengths is evident for significantly lower exciton densities (> 5 fold) with respect to the NB case.

We also performed density-dependent measurements at zero electric field, resulting in no sizeable energy shift of the main hIX peak and a less pronounced optical saturation (Supplementary Note 7). In particular, we observe bleaching of zero-field hIXs ($n_{hIX} > 10^{12}$ cm$^{-2}$ in narrowband) at densities about one order of magnitude higher with respect to high-field L-hIX and H-hIX. We attribute such strongly enhanced nonlinear response under a high electric field to the increased contribution of Coulomb exciton interactions. Thus, with an applied electric field we uncover strongly-interacting dipolar species with highly-nonlinear behaviour.

**Tunable hIX formation and relaxation dynamics**

We have performed ultrafast transient absorption measurements to unveil the time-dependent interactions and dynamics of electrically-tunable dipolar hIXs in our platform. Figures 3a-b show the 2D maps of the differential reflectivity ($\Delta R/R$) spectra as a function of pump-probe delay $\tau$ and probe photon energy in the cases of zero field and high field ($E_z \simeq 230$ mV/nm), respectively. In this experiment, the ~100 fs pump pulses were applied resonantly with $X_A$ (FWHM = 10 nm). Previous studies on monolayer TMDCs have shown that excitations resonant with $X_A$ induce an instantaneous build-up of the $X_B$ transient signal due to intravalley exchange interactions[48,49]. Thus, in MoS$_2$ bilayers, where the interlayer transition and $X_B$ are strongly hybridized, a sizeable hIX population is expected upon resonant photoexcitation of $X_A$. We employ this excitation scheme in order to exclude the role of exciton cascade effects, allowing us to decouple the excited ($X_A$) and hybridized species ($X_B$ and hIX), which remain connected through intravalley mixing processes. Further discussion on different photoexcitation schemes can be found in Supplementary Note 13.

In the 2D map of Figure 3a, positive and negative signals appear immediately after time zero ($\tau = 0^+$). Such features can be attributed to different pump-induced modifications of the excitonic spectrum (i.e. reduction of oscillator strength, broadening, and shift in energy). While purely symmetric derivative-shaped signals are a consequence of exciton line shifts due to exciton-exciton interactions[50] or bandgap renormalization[51], prominent positive peaks are mainly related to exciton absorption saturation due to the Pauli blocking effect[52]. Purely negative signals are generally attributed to photoinduced absorption[53], while exciton line broadening can lead to more complex shapes in the transient reflectivity spectra, but it becomes significant only at high pump fluences, since it is mainly caused by excitation-induced dephasing[52,54]. In our case, positive signals are always dominant in the transient reflectivity spectra at all the scanned time delays, pointing at the main role of Pauli blocking of the different excitonic interactions. We trace the intensity variation of such peaks to monitor the exciton population dynamics in the system. Supplementary Note 8 further expands on the extraction of exciton dynamics from transient absorption measurements[29,50,55].

At high $E_z$, with approximately 40 meV of H-L splitting in equilibrium conditions (Figure 3b), we observe positive signals associated with the photobleaching of both L-hIX and H-hIX (Figure 3c). Focusing on the energy range of hybrid excitons (Figure 3d), we observe a significant shift in the $\Delta R/R$ peak positions for both L and H species as a function of time. We track the temporal evolution of the photobleaching of the hIX peaks (Methods), revealing the prominent shift of the hIX energies on short timescales (< 0.3 ps), followed by a slower recovery to their unperturbed energies within tens of picoseconds (Figure 3e). In Supplementary Note 14, by taking into account the presence of heating effects, causing a linear blueshift at long timescales, we extract comparable and opposite shifts related to excitonic interactions for L-hIX and H-hIX at high fields.

Figure 4a compares the intensity of the transient signal of the L-hIX and h-hIX peaks with the one of hIX at zero field. The signals are fitted by an exponential rise ($\tau_R$) followed by a bi-exponential decay ($\tau_1$ and $\tau_2$), as commonly observed for intralayer exciton dynamics in TMDCs[56] (Methods). We note that the energy shifts shown in Figure 3e follow the timescales of the formation and relaxation dynamics observed by the $\Delta R/R$ signal presented in Figure 4a, being caused by density-dependent interactions. Upon an external electric field, L-hIX displays a clearly delayed formation, as well as an increase in the slow decay component (Figure 4a). This is also observed comparing the L-hIX dynamics to the $X_A$ and $X_B$ ones, taken from the same measurement at high field (Figure 4b). On the other hand, the transient behavior of H-hIX shows much faster formation and decay times with respect to L-hIX, the former being more similar to the other exciton species.

The formation time of purely-interlayer excitons in type-II TMDC heterobilayers is extended with respect to intralayer species due to the type-II band offset $\Delta E$ combined with phonon-assisted scattering mechanisms[39,40,57]. However, the dependence of the formation time of an interlayer species with respect to a tunable band offset has not been experimentally reported. In our case, if we consider $MoS_2$ homobilayer hIXs at the K point, a wider band offset $\Delta E$ between the K valleys of the two layers due to an external electric field results in a slower and less efficient momentum transfer (see inset of Figure 4a). In a single-particle picture at a specific valley, the interlayer tunneling matrix element between the two layers depends on the spatial overlap between the wavefunctions of the involved states[57]. In $MoS_2$ bilayers, the measured Stark shift magnitude represents the field-dependent offset between the valence band maxima at the K points in the two layers[23,44], as schematically presented in Figure 1b. With an applied field such that $\delta E_{H-L} = 40$ meV, corresponding to a band shift of approximately $\Delta E = 20$ meV, the decrease in spatial overlap between the electron and hole wavefunctions gives rise to a lower momentum transfer efficiency, resulting in a buildup time delay for species with higher interlayer composition factors (i.e. lower hybridization with $X_B$). Thus, the tunneling rate of the hybridized holes in H-hIXs (L-hIXs) is expected to increase (decrease) relative to the zero-field hIX tunneling rate ($J_0$). Considering H and L species at non-zero field with rates $J_H$ and $J_L$, this results in $J_L < J_0 < J_H$ (Figure 1b). This explains the longer formation of L-hIX compared to H-hIX and the other exciton species of interest (Figure 4b).

Figure 4c compares the fitted formation times of $X_A$, $X_B$ L-hIX and H-hIX at high field with those of hIX at zero field. While the zero-field hIX formation ($\tau_R^{0-hIX}$) occurs with a slight delay with respect to $X_B$ due to interlayer hybridization, an increase (decrease) in such delay is measured for L-hIX (H-hIX) at a high field. The observed increase of $\tau_R^{L-hIX}$ with respect to the zero-field case, with a ratio $\tau_R^{L-hIX}/\tau_R^{0-hIX} \sim 2.6$, can be explained by the reduction in the hole tunneling rate caused by an increased band offset $\Delta E$ under an applied electric field. Furthermore, the number of scattering events needed for a complete exciton thermalization also increases with larger energy offsets. This leads to an increase in exciton-phonon scattering rates[39,40], and to longer L-hIX formation times, in agreement with our experimental findings. Meanwhile, we observe the opposite behavior for H-hIX (i.e. lower $\tau_R^{H-hIX}$ due to a decreased band offset), although its quantification is limited by our temporal resolution.

Regarding the decay dynamics of L-hIX, we observe an increase in both short ($\tau_1$) and long ($\tau_2$) decay times of L-hIX at high field. However, the wide uncertainty of $\tau_1$ for L-hIX at high field, partially caused by the dominance of the slow component, does not allow us to infer conclusions on the modulation of the fast decay (see Supplementary Note 8). On the other hand, both short and long decay times of H-hIX are comparable to the zero-field counterpart. Thus, we focus here on the modulation of the slow decay $\tau_2$, occurring in the tens of picosecond time-scale (Figure 4d). Based on previous theoretical work on 2L-MoS$_2$, hIXs are expected to have radiative lifetimes on the order of picoseconds[58], thus contributing to $\tau_2$. By employing a formalism based on an exciton-photon interaction Hamiltonian (Supplementary Note 9), we find that the radiative decay rate of momentum-direct L-hIX excitons $\Gamma_{rad}^{KK}$ decreases with respect to interlayer composition as $\Gamma_{rad}^{KK}(E_z) \propto |1 - C_{IX}^{KK}(E_z)|^2$. Thus, we estimate a field-dependent increase in radiative lifetime $\tau_{rad}^{L-hIX}(\delta E = 40 \text{ meV})/\tau_{rad}^{hIX}(0) \sim 1.7$, which is compatible with the experimentally observed increase in $\tau_2$ of about 2. We note that a significant contribution to this modulation likely arises from phonon-assisted radiative recombination, which still scales as $\tau_{rad}$ with respect to the applied electric field.

To understand the role of second-order effects in the hybrid exciton dynamics, we performed fluence-dependent pump-probe experiments on the gated MoS$_2$ bilayer, focusing on the dynamics of the L-hIX under high electric field. Both the buildup and the long decay times are not strongly modulated within our scanned fluence range (Supplementary Note 8). This suggests that the slow buildup time of L-hIX observed under high field is not related to fluence-dependent phonon scattering processes[59]. Furthermore, exciton-exciton annihilation (Auger recombination) can be neglected in the observed exciton decay dynamics within our range of fluences, as recently demonstrated for $MoS_2$ monolayers[60], but previously found in other material systems[52,61,62]. In summary, we mainly attribute the measured field-induced increase of $\tau_2^{L-hIX}$ with respect to H-hIX, $X_A$ and $X_B$ at high field and hIX at zero field, to the suppression of the L-hIX (KK) radiative decay rate with increasing field-dependent interlayer mixing $C_{IX}$.

Recent developments in the literature have revealed that electrostatic doping can also induce sizable L-H splitting due to a quasi-static random coupling between L-hIX and H-hIX[32]. With this regard, in Supplementary Note 12 we further investigate and discuss the doping-dependent dynamics of L-hIX, unveiling an increase in formation and long decay times for higher $\delta E_{H-L}$. Albeit outside the scope of the present work, these findings might spark further research on the tunability of many-body electron correlations in TMDCs.

## DISCUSSION

We unveiled the strong Coulomb interactions between dipolar momentum-direct layer-hybridized excitons in a dual-gated van der Waals TMDC homobilayer system using nonlinear reflectivity measurements. Opposite energy shifts were observed for the two dipolar exciton species corresponding to their different dipole orientation, together with actively enhanced nonlinearities compared to the zero-field case. We demonstrated electrical control on the hybridized hole tunneling rate, resulting in slower hIX formation, as well as on the decay of hybrid excitons due to the increase in their interlayer character (i.e. decrease in hybridization). The measured density-dependent shifts and the increase of hybrid exciton decay time with electric field are supported by a microscopic and material-specific theory with predictive capabilities.

Strong nonlinearities and long lifetimes are essential for observing correlated states of interlayer excitons[63,64], while a high oscillator strength unlocks the potential for polariton condensation[15]. Our work indicates a route forward to achieve these feats, since unveiling and characterizing the electrical tunability of momentum-direct strongly-interacting dipolar species with high nonlinearity represents an essential step towards tunable macroscopic quantum states of matter in TMDC bilayers. Furthermore, electrically-tunable highly nonlinear exciton interactions will be crucial for developing ultrafast electro-optical polariton switches[29].

## METHODS

### Device fabrication

The device consists of an hBN-encapsulated $MoS_2$ homobilayer with bottom and top graphene gates over a $SiO_2$/Si substrate with an oxide thickness of 270 nm. The heterostructure was fabricated by a dry-transfer technique using polycarbonate (PC) membranes[65]. hBN and $MoS_2$ (SPI Supplies) were exfoliated onto $SiO_2$ and PDMS (gelpak) substrates, respectively. The flakes were identified by optical contrast. The heterostack was made by picking up the flakes successively using the PC stamp, and then released onto the bottom graphene gate by progressive adhesion while increasing the temperature above 150°C. The PC stamp was cleaned by chlorophorm. The heterostructure was annealed in high-vacuum ($10^{-6}$ mbar) for six hours at a temperature of 340 °C. Electrical contacts were fabricated by electron-beam lithography and evaporation of Ti/Au (2nm/80nm) layers.

### Optical measurements

For all measurements, the sample is maintained in a closed-cycle helium cryostat at 8 K with electrical feedthrough connections to the graphene gates and the homobilayer flake. Electrostatic gating is achieved using a two-channel Keithley 2612B.

Transient reflectivity measurements

For the transient measurements, 100-fs pulses from an amplified Ti:Sapphire laser (800 nm) at 2 kHz repetition rate are used. The laser output is split into two beams. One beam drives a non-collinear optical parametric amplifier (NOPA) to generate tunable pump pulses. The other beam is focused into a sapphire plate to generate the broadband white light probe pulse. The pulses are collinearly combined and focused on the sample using an objective lens. The two beams are cross polarised and the pump pulse is filtered out in the detection path by a linear polariser. The differential reflectivity (ΔR/R) spectra are recorded as a function of time (τ) by controlling the delay between pump and probe pulses using a mechanical delay line. Specifically, the probe reflectivity spectrum with the pump on ($R_{\text{PumpOn}}$) at each delay is compared to a reference spectrum taken without the pump ($R_{\text{PumpOff}}$). These are used to calculate. $\Delta R/R = (R_{\text{PumpOn}} - R_{\text{PumpOff}})/R_{\text{PumpOff}}$. The maps shown in Figure 3 are the ΔR/R values at each delay. The $\Delta R/R$ signals of interest in our case are well described by an exponential rise followed by a bi-exponential decay[56]:

$$\frac{\Delta R}{R}(t) = \text{IRF}(t) \otimes \left[\left(1 - e^{-\frac{t}{\tau_R}}\right) \cdot H(t) \cdot \left(A_1 e^{-\frac{t}{\tau_1}} + A_2 e^{-\frac{t}{\tau_2}}\right)\right] \qquad (2)$$

where IRF is the Gaussian instrument response function, $\tau_R$ is the rise time, $H$ is a Heaviside function, $\tau_1$ and $\tau_2$ are the fast and slow decay times. The fitting model is further discussed in Supplementary Note 11. The evolution of the H and L peaks in time (Figure 3) is fitted by a double-Lorentzian model.

Static reflectivity measurements

For the static reflectivity measurements (Figure 2), only the supercontinuum white light pulses (generated as described above) are used to excite the sample. The incident power and bandwidth are chosen with a subsequent use of variable filters.

**Microscopic many-particle theory**

To study layer-hybridized exciton states in TMDC bilayers, we derived a many-body Hamiltonian in a hybrid exciton basis that contained a kinetic part, the exciton–photon interaction and the exciton-exciton interaction relevant at elevated electron-hole densities. By solving the bilayer Wannier equation we obtained access to pure intra- and interlayer exciton states. By taking these states together with material-specific tunneling parameters (obtained by DFT calculations[66]) as input for a hybrid exciton eigenvalue problem we obtained the hybrid exciton landscape of spin- and momentum-bright exciton species. We included an out-of-plane electric field in our calculations by considering the quantum-confined Stark effect leading to energy shifts of interlayer resonances[42] (Supplementary Note 2). The hybrid exciton eigenstates were used to compute density-dependent energy renormalizations of hybrid excitons obtained from the Heisenberg equation of motion (Supplementary Note 4), as well as input for the hybrid exciton-photon interaction and the electric-field-dependent radiative recombination rates of hybrid excitons (Supplementary Note 9).

## DATA AVAILABILITY

The data that support the findings of this study are available on Zenodo at doi:10.5281/zenodo.17201075.

## ACKNOWLEDGEMENTS

We acknowledge fruitful discussions with Joakim Hagel (Chalmers University of Technology). We acknowledge the help of Z. Benes (EPFL Center of MicroNanoTechnology (CMI)) with electron-beam lithography. This project has received funding from the European Union's Horizon 2020 research and innovation programme under grant agreement No 956813 (Marie Curie Sklodowska ITN network "2-Exciting"). This work was financially supported by the Swiss National Science Foundation (grant no. 215089). AG, CL, SDC, CS, and GC acknowledge funding from the European Horizon EIC Pathfinder Open programme under grant agreement no. 101130384 (QUONDENSATE). This work reflects only authors' view and the European Commission is not responsible for any use that may be made of the information it contains. AG, SDC, and GC acknowledge financial support by the European Union's NextGenerationEU Programme with the I-PHOQS Infrastructure (IR0000016, ID D2B8D520, CUP B53C22001750006) "Integrated Infrastructure Initiative in Photonic and Quantum Sciences". The Marburg group acknowledges funding from the Deutsche Forschungsgemeinschaft (DFG, German Research Foundation) via SFB 1083 (project B9) as well as regular DFG project 512604469. K.W. and T.T. acknowledge support from JSPS KAKENHI (Grant Numbers 19H05790, 20H00354 and 21H05233).

## AUTHOR CONTRIBUTIONS

A.K., S.D.C., A.G., and C.G. initiated the project. A.K., S.D.C, A.G., and G.C. supervised the project. E.L. fabricated the device. E.L., C.L., and I.L. performed the optical measurements supervised by A.G. and they all analysed the data. A.G. and C.S. optimized the gate-tunable pump-probe microscopy setup. K.W. and T.T. grew the h-BN crystals. D.E, S.B, R. P-C and E.M developed the microscopic model. E.L, C.L., A.G. and A.K. wrote the manuscript with contributions from all authors.

**COMPETING INTERESTS**

The authors declare no competing financial interests.

**FIGURES**

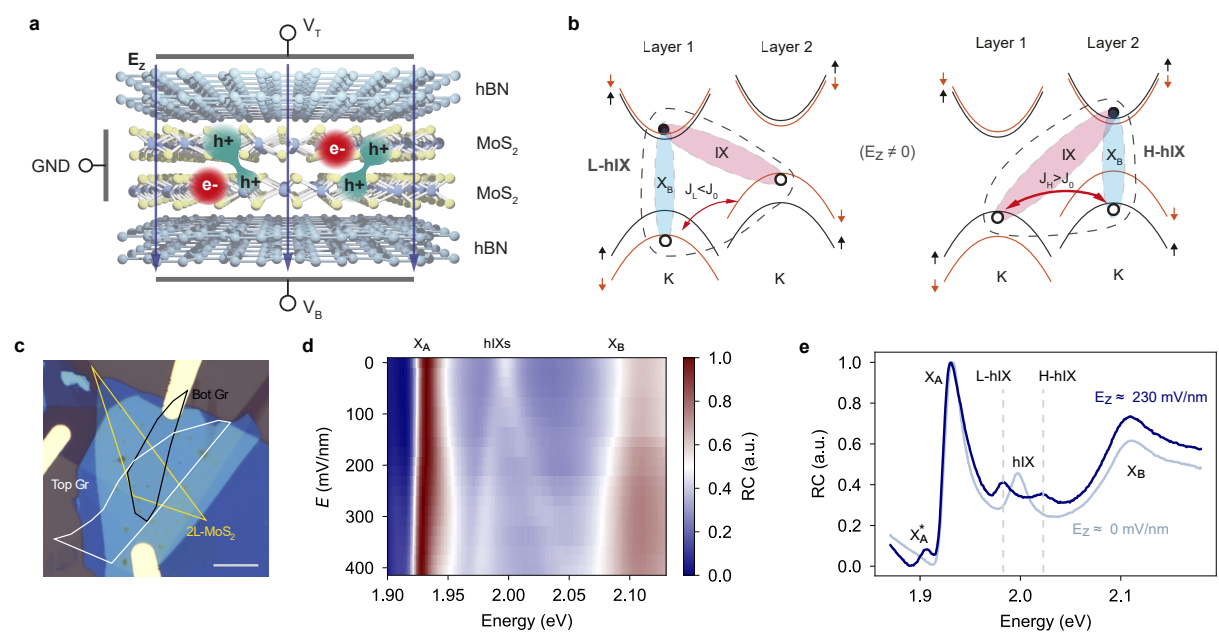

**Figure 1. Layer-hybridized excitonic transitions in a dual-gated MoS₂ homobilayer. a,** Schematic of the device structure, showing the hBN-encapsulated 2H-MoS$_2$ homobilayer with bottom and top graphene gates. **b,** Band diagram of the low-energy (L-hIX, left) and high-energy (H-hIX, right) hybrid excitons in a MoS$_2$ homobilayer under an applied electric field $E_Z \neq 0$. The holes of the interlayer species are strongly hybridized with the B excitons and delocalized in space between the two layers. With an applied vertical electric field, the tunnelling strengths $J_L$ of L-hIX ($J_H$ of H-hIX) are decreased (increased) with respect to the zero-field hIXs ($J_0$). **c,** Optical micrograph of our device, with coloured lines highlighting the MoS$_2$ homobilayer (yellow), as well as the bottom graphene (black) and the top graphene flakes (white). Scale bar: 10 μm. **d,** Reflectance contrast spectra as a function of the applied vertical electric field. The electric field $E_z$ is obtained as described in the Methods section. The hIX degeneracy is lifted upon application of an electric field, showing dipolar hybrid species with opposite orientations, as in (b). **e,** Reflectance contrast spectra from (d) at electric fields of 0 mV/nm (light blue) and 230 mV/nm (dark blue) with white light continuum

pulses and a fluence of 0.1 µJ cm$^{-2}$. All the main excitonic transitions of interest are highlighted at their respective energy positions.

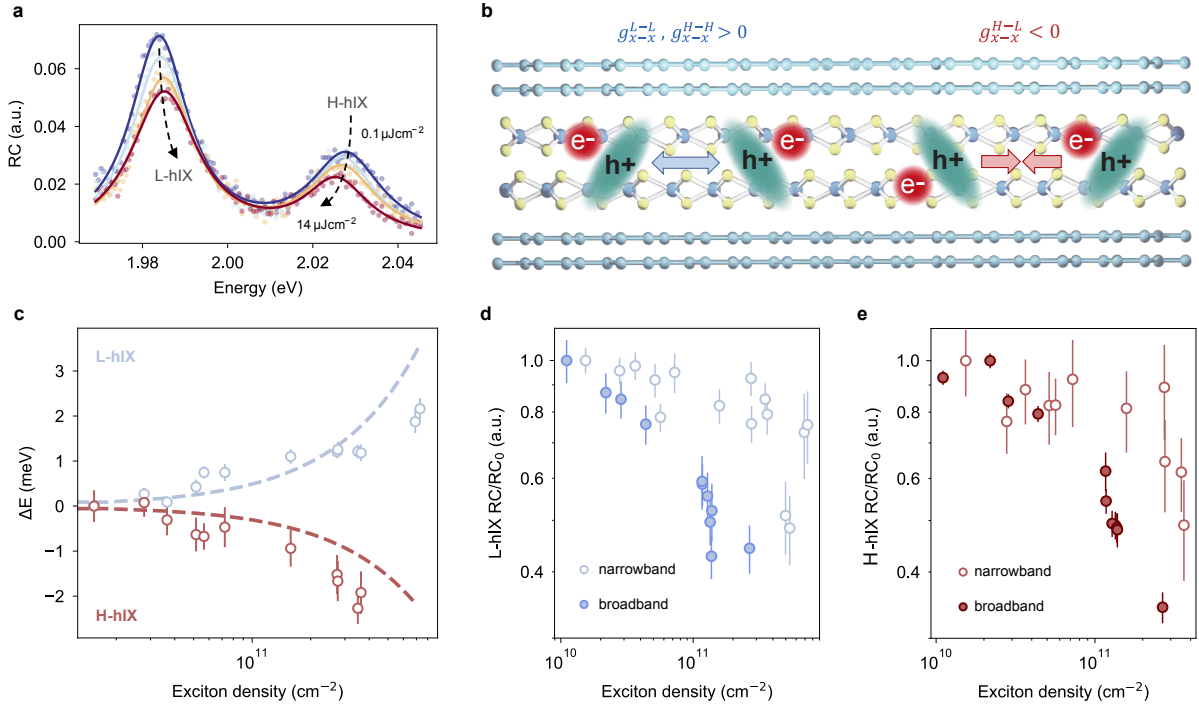

**Figure 2. Field-tunable interacting dipolar hybrid excitons. a,** Reflectance contrast spectra as a function of excitation density at a fixed applied vertical electric field $E_z = 230$ mV/nm, corresponding to a low-density energy splitting between hIX species of approximately 40 meV. The sample was excited with ultrashort (100 fs) pulses, filtered in energy to selectively excite the hIX species in a narrowband configuration (FWHM = 100 meV). The excitation fluences range from 0.1 to 14 µJ cm$^{-2}$ increasing magnitude following the black dashed lines. The superimposed fits to the data (smooth lines) are obtained using a Lorentzian model. The energy shift of the two exciton species is observed upon increasing the pulse fluence. The L-hIX and H-hIX energy peaks undergo a blueshift and redshift, respectively. A comparable behaviour is obtained in the broadband configuration (FWHM = 200 meV), as shown in Supplementary Figure 6. **b,** Illustration of the dipolar interactions between hIXs in a MoS$_2$ homobilayer with an applied electric field. Same-species interactions (L-L and H-H) result in a density-dependent blueshift, while attractive interactions between opposite species (L-H) result in a net density-dependent redshift. **c,** Density-dependent energy shifts for L-hIX (blue) and H-hIX (red) in the narrowband configuration, compared with the corresponding calculations from microscopic theory (dashed lines), with $n_L \approx 2\,n_H$ as a best-fit approach (Supplementary Note 4). The x-axis is displayed in logarithmic scale to cover the entire range of the applied pump fluences. The hybrid exciton densities are calculated taking into account the pump fluences and the measured RC spectra, as described in the Methods section and Supplementary Note 5. **d-e,** Normalized integrated reflectance contrast of L-hIX (b) and H-hIX (c) with respect to their corresponding population densities, for narroband (empty markers) and broadband (full markers) pump configurations. A subset of the spectra obtained by narrowband excitations is shown in (a), while the respective broadband cases are displayed in Supplementary Figure 6. All error bars in (c-e) represent the standard deviations of the quantities of interest extracted from the double-Lorentzian fits in (a).

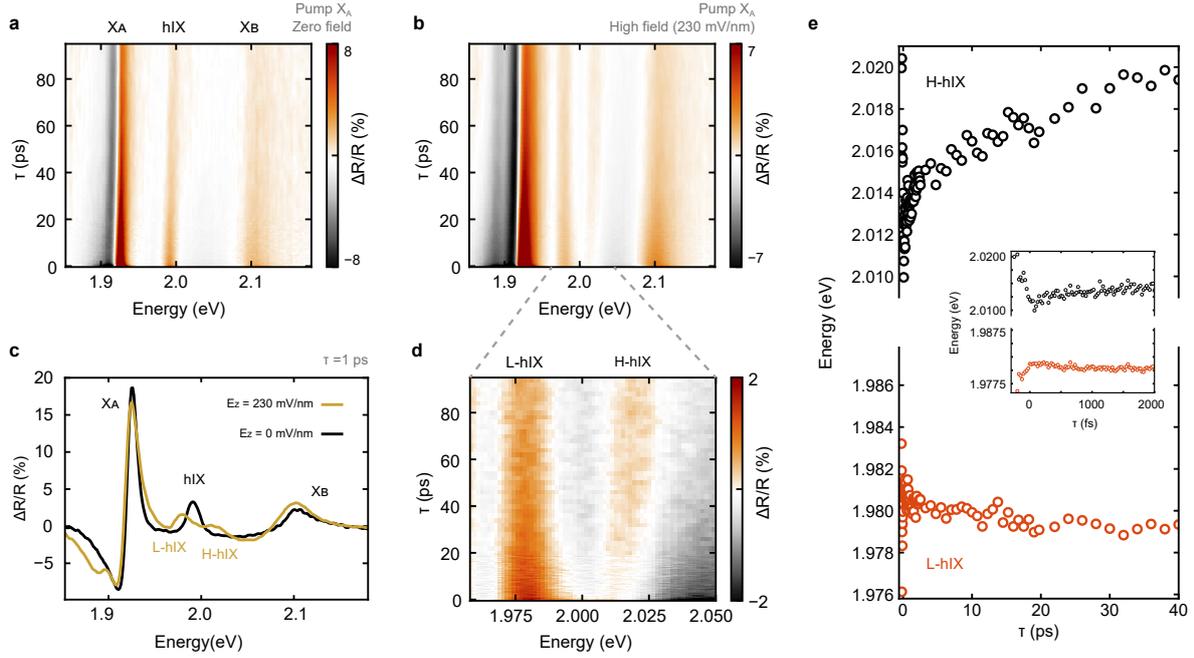

**Figure 3. Time-domain observation of layer-hybridized exciton interactions. a-b,** Transient differential reflectance ($\Delta R/R$) spectra as a function of the pump-probe delay $\tau$, obtained by exciting the structure with a pump pulse (~100 fs) resonant to $X_A$ and probing with a broad white-light continuum pulse at electric fields $E_z = 0$ mV/nm (a) and $E_z = 230$ mV/nm (b). **c,** $\Delta R/R$ traces extracted from (a) and (b) at a pump-probe delay $\tau = 1$ ps. All transitions of interest are highlighted. **d,** $\Delta R/R$ traces in time in the hIX energy range. From these signals, we observe a change in time of the H-L splitting $\delta E_{H-L}$ on the picosecond timescale. We note that the full $\Delta R/R$ intensity scale from (b) was reduced in (d) in order to visualize the hIXs signals. **e,** In order to quantify and understand the observed energy shifts, we fitted the $\Delta R/R$ signals in (e) with a double Lorentzian in order to track the temporal evolution of the L-hIX and H-hIX peaks. We observe two regimes, for $\tau < 0.3$ ps and $\tau > 0.3$ ps. Right after pump excitation, the build-up of the hIX density results in an increase of the hybrid exciton dipolar interactions, inducing a transient reduction of $\delta E_{H-L}$. This is followed by exciton depopulation processes decreasing the L-hIX and H-hIX densities within tens of picoseconds, leading to an opposite energy shift for both peaks. Since the energy shifts are directly proportional to the hybrid exciton density $n_{hIX}$, the energy dependent shifts follow the trend of the corresponding $\Delta R/R$ intensity traces. Inset: zoom-in on the first 2 ps of the time-dependent L and H energy peak shifts.

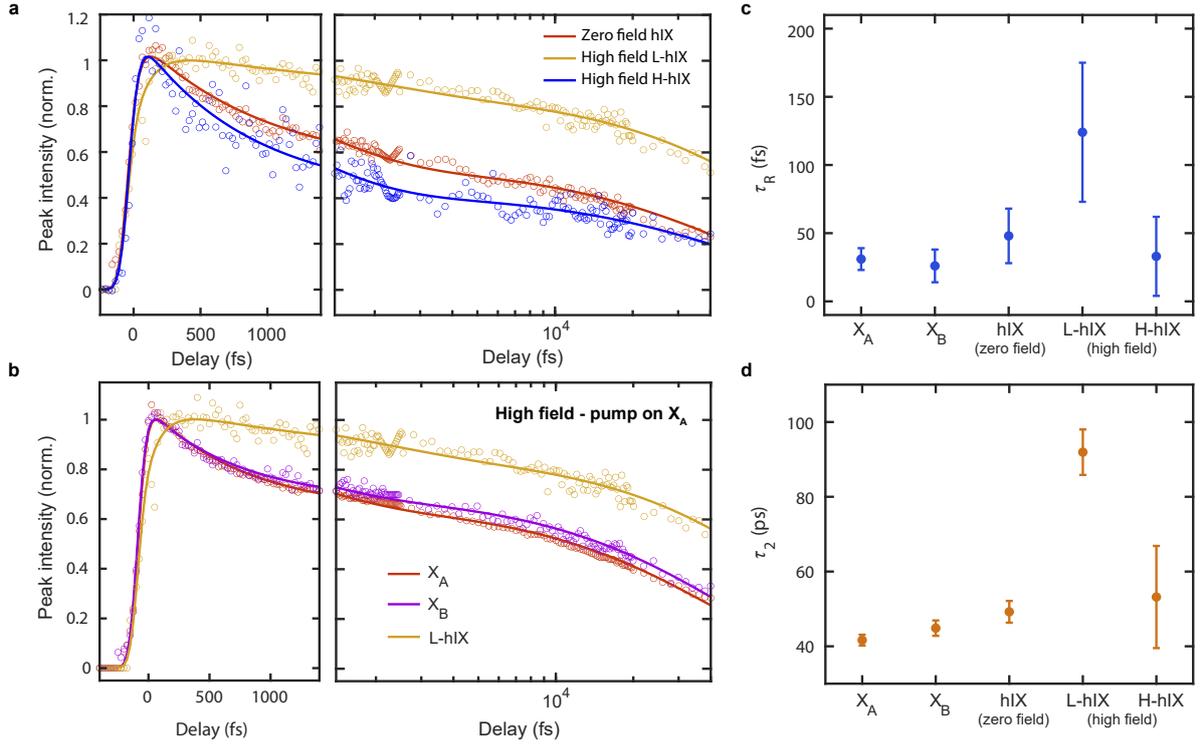

**Figure 4. Electrical control of ultrafast hIX dynamics. a,** Transient intensity of the hIX peak extracted from the $\Delta R/R$ spectra at zero electric field (red) compared with that extracted at high field from the L-hIX (yellow) and H-hIX (blue) peaks with $\delta E(0) = 40$ meV at a pump fluence of 12.4 µJ cm$^{-2}$. The transient intensities were obtained by tracking the peak energy positions based on their time-dependent energy shifts, as described in the Methods section. The solid lines are fits to the data obtained using the model described in the Methods section (Equation 2). **b,** $\Delta R/R$ intensity of $X_A$ (red), $X_B$ (purple) and L-hIX (yellow) at $E_z = 230$ mV/nm, with respective fits. **c-d,** Formation time $\tau_R$ and long decay time $\tau_2$ extracted from the fits in (a-b). For hIX at zero field, we obtain $\tau_R^{0-hIX} = 48 \pm 20$ fs and $\tau_2^{0-hIX} = 49 \pm 3$ ps. At high field, we extract $\tau_R^A = 31 \pm 8$ fs and $\tau_2^A = 41 \pm 1$ ps for $X_A$, $\tau_R^B = 26 \pm 12$ fs and $\tau_2^B = 45 \pm 2$ ps for $X_B$, $\tau_R^{L-hIX} = 124 \pm 51$ fs and $\tau_2^{L-hIX} = 92 \pm 6$ ps for L-hIX, as well as $\tau_R^{H-hIX} = 33 \pm 29$ fs and $\tau_2^{H-hIX} = 53 \pm 14$ ps for H-hIX. $X_A$ and $X_B$ exhibit comparable dynamics at zero field. In Supplementary Note 8, we further compare all short decay $\tau_1$ values for the excitonic species of interest. All error bars in (c-d) represent the standard deviations of the extracted quantities of interest.

23